\documentclass[10pt,journal,compsoc]{IEEEtran}
\usepackage{array}
\newcolumntype{M}[1]{>{\centering\arraybackslash}m{#1}}

\usepackage{amssymb}
\usepackage{tikz}
\usepackage{comment}
\usepackage{pifont} 

\usepackage{amsmath}
\usepackage{amssymb}
\usepackage{amsthm}
\usepackage{amsmath}
\usepackage{algorithm,algpseudocode}
\usepackage{stfloats} 
\usepackage{hyperref}

\algrenewcommand\algorithmicrequire{\textbf{Input:}}
\algrenewcommand\algorithmicensure{\textbf{Output:}}
\usepackage{cite}
\usepackage{amsmath,amssymb,amsfonts}
\usepackage{graphicx}
\usepackage{textcomp}
\usepackage{xcolor}
\usepackage{mdframed}
\usepackage{caption}
\usepackage{xspace}
\usepackage{dsfont}
\usepackage{booktabs, makecell, tabularx}
\usepackage{comment}
\usepackage{subcaption}
\usepackage{capt-of}

\theoremstyle{plain}

\theoremstyle{definition}

\usepackage{subcaption}
\usepackage[T1, OT1]{fontenc}

\usepackage{xcolor}
\DeclareTextSymbolDefault{\dh}{T1}
\DeclareTextSymbolDefault{\TH}{T1}

\usepackage{multirow}

\usepackage{framed}

\ifCLASSOPTIONcompsoc

\fi
\ifCLASSINFOpdf
\else
\fi
\DeclareCaptionType[fileext=ext]{Protocol}[Protocol]
\DeclareCaptionType[within=none,fileext=ext]{Sub-Protocol-1}[Sub-Protocol 1.]
\DeclareCaptionType[within=none,fileext=ext]{Sub-Protocol-2}[Sub-Protocol 2.]
\DeclareCaptionType[fileext=ext]{Construction}[Construction]

\usepackage{stmaryrd}
\usepackage{amsmath}

\usepackage{amsthm}
\usepackage{framed}
\usepackage{pbox}

\usepackage{enumitem}

\begin{document}
\bstctlcite{BSTcontrol}

\title{Adaptive Honeypot Allocation in Multi‑Attacker Networks via Bayesian Stackelberg Games}

\author{Dongyoung Park, Gaby G. Dagher}

\IEEEtitleabstractindextext{
\begin{abstract}

Defending against sophisticated cyber threats demands strategic allocation of limited security resources across complex network infrastructures. When the defender has limited defensive resources, the complexity of coordinating honeypot placements across hundreds of nodes grows exponentially. 
In this paper, we present a multi-attacker Bayesian Stackelberg framework modeling concurrent adversaries attempting to breach a directed network of system components. Our approach uniquely characterizes each adversary through distinct target preferences, exploit capabilities, and associated costs, while enabling defenders to strategically deploy honeypots at critical network positions. By integrating a multi-follower Stackelberg formulation with dynamic Bayesian belief updates, our framework allows defenders to continuously refine their understanding of attacker intentions based on actions detected through Intrusion Detection Systems (IDS). Experimental results show that the proposed method prevents attack success within a few rounds and scales well up to networks of 500 nodes with more than 1,500 edges, maintaining tractable run times. 
\end{abstract}

}

\maketitle

\IEEEdisplaynontitleabstractindextext

%
\IEEEpeerreviewmaketitle


\IEEEraisesectionheading{\section{Introduction}\label{sec:introduction}}
In an increasingly digitalized world, cybersecurity has become a critical concern for safeguarding sensitive information and infrastructure. The rise of sophisticated cyberattacks demands innovative and adaptive defense mechanisms to protect network systems from adversarial threats. Among these, the challenge of predicting attackers' targets and devising optimal defense strategies remains at the forefront of cybersecurity research.

Game theory, a mathematical framework for analyzing strategic interactions between rational decision-makers, offers a powerful tool for modeling and mitigating cyber threats~\cite{carroll2011game}\cite{li2018targets}\cite{xu2005feasibility}. By conceptualizing the interactions between defenders and attackers as a game, it becomes possible to anticipate adversarial behaviors and design optimal defensive strategies. In particular, game-theoretic models enable defenders to allocate limited security resources efficiently and adaptively in response to threats~\cite{ attiah2018game}\cite{gibbons1992game}\cite{shiva2010game}.

Traditional security models often adopt a zero-sum game perspective, where one player's gain is inherently the other's loss~\cite{fudenberg1991game}. This assumption simplifies analysis but does not accurately reflect most cybersecurity scenarios. In reality, cyber defense and attack scenarios are more accurately described as non-zero-sum games, where the outcomes are not strictly opposing. Both attackers and defenders may incur varying costs and gains depending on their strategies and the environment. In non-zero-sum games, the defender's success in mitigating an attack does not necessarily equate to a complete loss for the attacker, and vice versa~\cite{attiah2018game}\cite{ferdowsi2017colonel}\cite{ jahan2020non}. This dynamic interplay requires more nuanced modeling to capture the partial successes and failures of both parties. For instance, an attacker might gain partial system access without full control, and a defender might mitigate damage without entirely preventing the attack. Therefore, employing non-zero-sum game models allows for more realistic and effective security strategies.

One of the most effective game-theoretic frameworks for cybersecurity is the Bayesian Stackelberg Game. In this leader-follower game model, the defender (leader) commits to a strategy first, and the attacker (follower) observes and responds accordingly~\cite{fudenberg1991game}\cite{paruchuri2007efficient}. This sequential interaction mirrors real-world security scenarios, where defenders must preemptively deploy resources while anticipating attacker reactions. The Bayesian Stackelberg Game is particularly useful for modeling scenarios involving resource allocation~\cite{brown2006defending}\cite{paruchuri2007efficient}\cite{ cardinal2005pricing}, such as deploying intrusion detection systems or honeypots in a network. The defender aims to maximize security effectiveness by strategically placing limited defensive resources~\cite{wahab2017know}~\cite{wahab2019resource}, while the attacker seeks to exploit the defender's configuration. However, finding the optimal solution in a Bayesian Stackelberg Game with one defender and multiple attackers is NP-hard~\cite{conitzer2006computing}. This computational complexity makes it extremely challenging to derive optimal strategies in such games. To address this, a Mixed-Integer Linear Programming (MILP) model allows us to compute the optimal defense strategy while significantly reducing the time complexity~\cite{sandholm2005mixed}. The MILP formulation efficiently handles the combinatorial nature of honeypot placement and attacker responses, providing a scalable and practical solution for real-world cybersecurity defense.

Modern networks often face multiple concurrent attackers, each potentially aiming at different targets or exploiting different vulnerabilities~\cite{nvd2025}. This multi-attacker environment significantly amplifies the complexity of defense. First, defenders must identify and prioritize incoming threats across various parts of the network with limited security resources. Second, attackers may coordinate indirectly or directly, forcing defenders to consider the possibility of attacks that exploit complementary vulnerabilities. Third, each attacker can exhibit distinct tactics, techniques, and procedures based on resources and objectives, further complicating accurate prediction of attack paths. These factors together make it exceedingly difficult to design a unified defensive strategy that accounts for all potential attacker types under real-world constraints.

To address this challenge, we propose an adaptive defense model using a Bayesian Stackelberg game with multiple attackers. Our model leverages dynamic belief updates and a strategic deployment of honeypots to guide attackers into revealing their intentions early. By partitioning the network into distance layers and modeling each subgame via a tractable Mixed-Integer Programming (MIP) approach, we derive a global defensive strategy that balances complexity and optimality. In each layer, the defender pre-commits to a mixed strategy, and attackers, observing this configuration, choose their best responses. We then combine these subgame solutions by backward induction, producing a proactive, adaptive defense that confronts the inherent difficulty of multiple attackers—each with unique attack graphs and objectives.

By strategically guiding attackers into revealing their intentions, the proposed approach allows the defender to find and implement the most effective defensive strategy. This proactive and adaptive approach significantly improves the defender’s ability to anticipate and disrupt attacks, ultimately enhancing the network's resilience against sophisticated and evolving cyber threats.

\subsection{Contribution}

The main contributions of our paper can be summarized as follows:

\begin{itemize}

\item We propose a novel Bayesian Stackelberg framework that explicitly models the strategic interactions between a single defender and multiple concurrent attackers on an attack graph. This formulation accurately captures the sequential decision-making and asymmetric information structures characteristic of realistic cybersecurity scenarios.

\item We introduce an adaptive defense mechanism driven by network graph modeling, which leverages a MILP formulation to optimally deploy honeypots. This approach systematically assesses risks across different network layers, allocates limited honeypots to disrupt attackers' most viable penetration paths, and significantly reduces the attackers' overall success probability.

\item We develop a Bayesian updating method that enables the defender to dynamically refine their beliefs about attackers’ targets and strategies based on observations. As the defender progressively accumulates information, their defense strategies become increasingly targeted, allowing proactive adjustments to effectively distort attacker routes and protect vulnerable nodes.

\item We evaluate the proposed approach through extensive experiments on networks of varying sizes and multiple attacker-skill scenarios. Experimental data show how dynamic honeypot placement rapidly decreases attack success rates to zero within a few rounds. Comparisons against greedy and random allocations reveal clear performance advantages. The runtime analysis demonstrates the framework’s feasibility for larger and more complex networks.

\end{itemize}

The rest of this paper is organized as follows. Section~\ref{sec:Related Work} reviews existing literature and relevant studies on Stackelberg security games and honeypot deployment strategies. Section~\ref{sec:System Model} describes our system model, providing detailed explanations of the network environment and adversary characteristics. Section~\ref{sec:Proposed Game_Theoretic_Framework} presents the proposed Bayesian Stackelberg formulation, focusing on Bayesian belief updates and optimal path selection for attackers. Section~\ref{sec:Implementation} outlines the implementation details of our proposed framework. Section~\ref{sec:case-study} provides a detailed case study demonstrating our method on a practical network scenario. Section~\ref{sec:Experimental_Evaluation} presents experimental results, analyzing sensitivity and scalability performance under varying network conditions. Finally, Section~\ref{sec:conclusion} summarizes key findings and suggests directions for future research.

\section{Related Work}\label{sec:Related Work}


Early research adopted classical strategic‐form games—predominantly zero-sum and non-zero-sum—to capture attacker–defender interactions, producing static defense recommendations that presuppose perfect information~\cite{carroll2011game,li2018targets,xu2005feasibility}~\cite{clark2012deceptive}.
While foundational, these one-shot formulations fail to reflect the sequential, information‐imperfect nature of real intrusions.


Stackelberg security games (SSGs) elevated realism by letting a defender (leader) pre-commit to a randomized allocation that attackers (followers) observe before responding. Pioneering deployments—ARMOR for airport checkpoints and IRIS for flight scheduling—demonstrated operational viability~\cite{tambe2011security}.
Algorithmic advances such as DOBSS~\cite{paruchuri2008playing} and ASP~\cite{korzhyk2011solving} exploit hierarchical structure to compute equilibria over vast action spaces.

A key theoretical concern is whether Stackelberg solutions remain effective when attackers do \emph{not} observe the defender’s mixed strategy. Yin \emph{et al.} prove that, for a broad class of SSGs, the defender’s Stackelberg strategy is also a Nash-equilibrium strategy and that Nash profiles are interchangeable, eliminating equilibrium-selection ambiguity for simultaneous-move variants~\cite{yin2010stackelberg}. Their analysis underpins many later works that assume commitment still yields optimality even with observation uncertainty.


Real networks involve incomplete information about attacker skill, intent, or resources. Bayesian Stackelberg games capture this uncertainty by defining a prior distribution over attacker types~\cite{harsanyi1967games}~\cite{wahab2019resource}.
Subsequent studies incorporate online belief revision, where partial observations refine type probabilities over repeated plays, enhancing defensive adaptability~\cite{yin2010stackelberg}.


Most foundational SSGs focus on a single follower, yet many infrastructures face concurrent adversaries. Multi-attacker extensions analyze how simultaneous incursions alter optimal allocations and utility trade-offs, often leveraging decomposition to contain complexity~\cite{basilico2017coordinating}. Our work builds on this line by modeling heterogeneous attackers who share the network but act non-cooperatively with respect to the defender.


Attack campaigns are typically stage-wise. Multi-stage games, including Markov and dynamic Stackelberg formulations, track an adversary’s progression across layered defenses~\cite{pawlick2019game}~\cite{durkota2015optimal}. Such models require backward‐induction or state-space expansion techniques that compound computational cost.


Deception defenses deploy honeypots or honey‐files to divert, observe, and sometimes block attackers~\cite{spitzner2003honeypots}. Game‐theoretic analyses weigh deployment cost against detection gain~\cite{pawlick2021game}, typically under resourcing limits that cap the number of decoys.

Despite these advancements, challenges remain in scaling such models to large networks and multiple attackers, dealing with uncertainties in attacker information, and designing efficient algorithms for computing optimal defense strategies. Integrating Bayesian Stackelberg games with adaptive honeypot allocation offers a promising direction to tackle these issues by enabling defenders to learn and adapt to attacker behaviors while accounting for incomplete and uncertain information. The resulting strategies can improve network resilience by maximizing the likelihood of attackers interacting with honeypots and minimizing the impact on the network.

\begin{table}
    \centering
    \caption{Notation Summary}
    \includegraphics[width= 1 \linewidth]{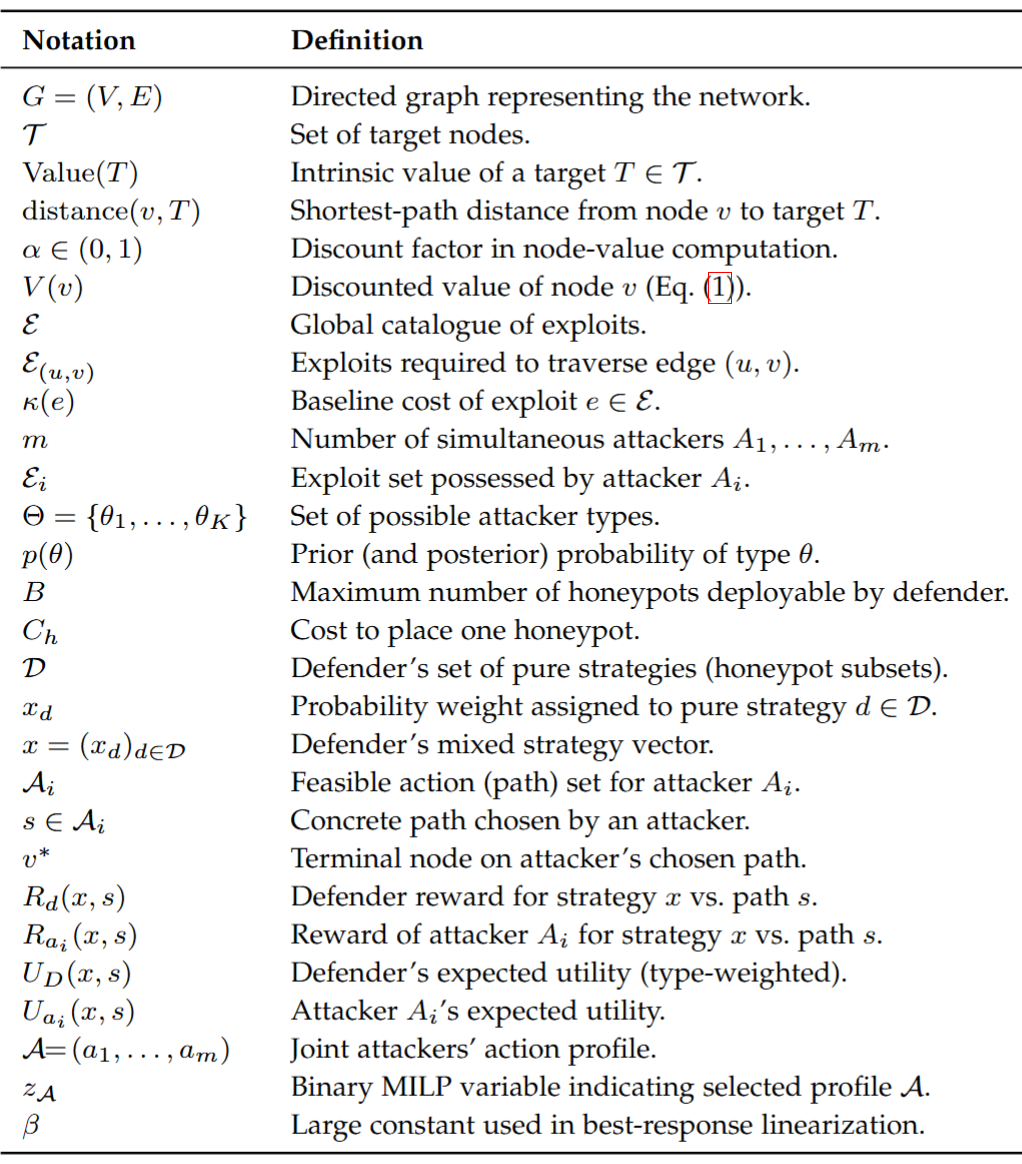}
    \label{table:alpha_impact}
\end{table}


\section{System Model}\label{sec:System Model}

This section introduces the fundamental concepts and notations necessary for modeling our multi-attacker cybersecurity scenario. We begin by defining the network environment and core elements of a Bayesian Stackelberg Game.

\subsection{Network Environment}

We consider a directed graph \( G = (V, E) \) that represents a networked system. Each node \( v \in V \) corresponds to a system component (e.g., a server, workstation, or service), and each edge \( (u, v) \in E \) indicates a feasible transition from node \( u \) to node \( v \).  $\mathcal{T}$ is the set of target nodes of interest, each target $T \in \mathcal{T}$ having an intrinsic value $\mathrm{Value}(T)$. For any node $v$, let $\mathrm{distance}(v,T)$ be the shortest-path distance from $v$ to $T$. The value of node \( v \) is defined as follows:

 \begin{equation}\label{eq:node_value}
 V(v) \;=\; \sum_{T \in \mathcal{T}} \Bigl(\alpha^{\mathrm{distance}(v,T)} \times \mathrm{Value}(T)\Bigr),
 \end{equation}
where $\alpha \in (0,1)$ is a discount factor. Nodes closer to a target retain higher value for the attacker.

Each edge \((u,v)\) typically requires at least one exploit from a set \(\mathcal{E}_{(u,v)}\) for an attacker to move from \( u \) to \( v \). Each exploit \( e \in \mathcal{E}_{(u,v)}\) has an associated cost \(\kappa(e)\). We assume a single defender who monitors the network via an Intrusion Detection System (IDS). The defender can deploy a limited number of honeypots on nodes to intercept adversaries and gather intelligence about their intentions. Each honeypot placement on node \( v \in V \) has a deployment cost \( C_h \). An attacker who enters a honeypot node risks being detected or misled, which provides the defender with new information for belief updates.
\subsection{Adversary Model}\label{sec:Adversary_Model}

\subsubsection{Multi-Attacker Setting and Attack Graphs}
We consider $m$ concurrent attackers $\{A_{1}, A_{2}, \ldots, A_{m}\}$ operating on the directed graph $G=(V,E)$. Each attacker $A_i$ possesses complete knowledge of the network topology and vulnerabilities but remains uncertain about the exact placement of defender-deployed honeypots. Every attacker aims to reach the target node $T_{i}\in V$ while incurring minimal exploitation costs. 

For each attacker $A_i$, a distinct attack graph $G_i \subseteq G$ captures the set of edges \((u,v)\) that attacker $A_i$ can feasibly traverse. Formally, $G_i$ is defined based on the exploits $\mathcal{E}_i$ available to $A_i$, and for any directed edge $(u,v) \in G$, the edge is included in $G_i$ if and only if:
\[
   \mathcal{E}_{(u,v)} \,\cap\, \mathcal{E}_i \;\neq\;\varnothing,
\]
where $\mathcal{E}_{(u,v)}$ is the set of exploits required to move from node $u$ to node $v$, and $\mathcal{E}_i$ is the set of exploits that $A_i$ is capable of utilizing. Consequently, each $G_i$ summarizes the potential paths that attacker $A_i$ may follow to reach a target.

\subsubsection{Attacker Capabilities and Objectives}
Each attacker $A_i$ is characterized by:
\begin{itemize}
    \item \textbf{Skill level and exploit set} $\mathcal{E}_i$: Higher-skill attackers may have advanced or more costly exploits, while lower-skill attackers may be restricted to a cheaper subset.
    \item \textbf{Target preference} $T_i \in V$: The designated target node that attacker $A_i$ desires to compromise.
    \item \textbf{Cost structure}: Moving along an edge \((u,v)\) requires at least one exploit $e \in \mathcal{E}_{(u,v)}$. If multiple exploits are feasible, $A_i$ selects the cheapest among them:
    \[
      \text{EdgeCost}_{a_i}(u,v)
      \;=\;
      \min_{\,e\,\in\, \mathcal{E}_{(u,v)} \,\cap\, \mathcal{E}_i}\kappa(e),
    \]
    where $\kappa(e)$ denotes the baseline cost of exploit $e$.
\end{itemize}
Given these elements, each attacker $A_i$ chooses a path $s$ (a sequence of edges) to reach its designated target $T_i$ while minimizing the sum of exploit costs.

\section{Proposed Game-Theoretic Framework}\label{sec:Proposed Game_Theoretic_Framework}

In this section, we propose a non-zero-sum, non-cooperative Bayesian Stackelberg game model to represent strategic interactions among attackers and a defender. We assume multiple attacker types, introducing uncertainty that the defender faces and must handle strategically. Let 
\[
\Theta \;=\; \{\theta_{1}, \theta_{2}, \ldots, \theta_{K}\}
\]
be the set of possible attacker types, each characterized by properties such as skill level, exploit set, target preferences, or tactics. A probability distribution \(p(\theta_{k})\) over these types encodes the defender’s uncertainty.
\subsection{Bayesian Stackelberg Game Formulation}\label{subsec:Bayesian_Stackelberg_Game_Formulation}

The key steps in this formulation are:

\begin{enumerate}
    \item \textbf{Leader's Commitment:} 
    The defender commits to a mixed strategy for placing honeypots across the network. The defender's uncertainty over attacker types \(\theta \in \Theta\) is represented by the prior probabilities \(p(\theta)\).

    \item \textbf{Followers' Observation and Response:} 
    Each attacker partially observes the defender's strategy and chooses a best-response path or exploit method that maximizes its own expected utility. Different attacker types may have different objectives, exploit costs, or target values.

    \item \textbf{Bayesian Updating:} 
    When the defender observes specific attacker actions, these observations inform Bayesian belief updates about the attackers' types. Future defensive actions are then adjusted accordingly.

    \item \textbf{Equilibrium Outcome:} 
    The solution concept is a Stackelberg equilibrium, where the defender's strategy anticipates the attackers' best responses under the type. 
\end{enumerate}

This structure ensures that the defender optimizes its expected utility with respect to the probabilities of different attacker types and that each attacker, given the defender’s chosen mixed strategy, has no incentive to deviate from its best response. 

\subsection{Defender's and Attacker's Actions}

\subsubsection{Defender Actions}

The defender’s actions involve deciding which nodes to place honeypots on:

\begin{itemize}
    \item \textbf{Pure strategy}: A concrete subset of nodes \(x\subseteq V\) on which honeypots are installed, satisfying
    \begin{equation}\label{eq:capConstraint}
      \sum_{n \in x} 1 \;\;\le\;\; B.
    \end{equation}

    \item \textbf{Mixed strategy}: A probability distribution over all feasible pure strategies \(\{x_i\}\), ensuring
    \begin{equation}\label{eq:mixedStrategy}
      \sum_i p_i \;=\; 1,
      \quad
      p_i \;\ge\; 0.
    \end{equation}
\end{itemize}

\subsubsection{Attacker Actions}

Attackers select a path to reach a high-value target node. In the multi-stage version, each attacker starts from an entry node and proceeds step by step. If an attacker lands on a honeypot node, it is immediately detected, typically yielding a negative reward for the attacker from that point.

\subsection{Reward Function}\label{sec:Reward_Function}

\subsubsection{Defender Reward \texorpdfstring{$(R_d$)}{:}}

A common modeling approach is to define the defender's reward in terms of a node’s value and the cost of deploying honeypots. Let \(V(v)\) be the value of node \(v\), and let \(C_h\) be the cost of placing a honeypot. If the attacker’s path \(s\) includes a node \(v\) protected by a honeypot, the defender gains \(V(v) - C_h\). Conversely, if the attacker successfully reaches a node \(v\) without encountering a honeypot, the defender suffers a loss \(-\,V(v)\). A simplified version of this can be written as:
\begin{equation}\label{eq:defenderReward}
R_d(x, s) \;=\;
\begin{cases}
    V(v) \;-\; C_h, & \text{if } A, D \rightarrow v,\\
    -\,V(v),        & \text{if } A \rightarrow v, D \not\rightarrow v,\\
    0,              & \text{if } A, D \not\rightarrow v.
\end{cases}
\end{equation}
Here, \(x\) denotes a pure defensive strategy (i.e., the set of nodes with honeypots installed). In practice, we often sum contributions over all relevant edges and nodes in the path or consider multiple stages of the attack. For example,
\[
R_d(x, s)
\;=\;
\sum_{\,e \in x} [-\,C_h]
\;+\;
\sum_{\substack{e \in x \\ e \in s}}
\bigl[V(v) - C_h\bigr]
\;+\;
\sum_{\substack{e \notin x \\ e \in s}}
\bigl[-\,V(v)\bigr],
\]
indicating that each honeypot placement costs \(-\,C_h\), while intercepting an attacker at a protected node \(v\) yields \(V(v) - C_h\). If the attacker bypasses the honeypot, the defender loses \(-\,V(v)\) on that node.

\subsubsection{Attacker Reward \texorpdfstring{$(R_{a_i})$}{:}}

In our model, each attacker \(A_i\) is characterized by a specific skill level, which determines the set of exploits \(\mathcal{E}_i\) that the attacker can use. For instance, a high-skill attacker may possess the full set of eight exploits, including advanced and expensive ones, while a low-skill attacker might have access only to some of the cheapest exploits. This distinction directly influences which edges the attacker can traverse and what cost it pays on each edge.


The total path cost is then the sum of these edge-level costs over the path \(s\). Then, a general form of the attacker’s reward is:

\begin{equation}\label{eq:attackerReward_revised}
R_{a_i}(x,s)=
\begin{cases}
    -E_{c_i}, &
    \text{if } A,D \rightarrow v,\\[4pt]
    V\!\bigl(v^{*}\bigr)-\displaystyle\sum_{(u,v)\in s}
       \min_{\substack{e\in\mathcal{E}_{(u,v)}\\ e\in\mathcal{E}_i}}
       \kappa(e), &
    \text{if } A\rightarrow v,\; D\not\!\rightarrow v.
\end{cases}
\end{equation}

\subsection{Bayesian Belief}

The defender updates its belief about attacker types at each stage based on observed attacker actions (e.g., probing nodes, exploit usage). This update is performed via Bayes' rule as follows:

\begin{enumerate}
    \item \textbf{Initial Belief}: \(\Pr(\theta)\) for all \(\theta \in \Theta\).
    \item \textbf{Observation}: At stage \(t\), the defender observes event \(E_t\), which may include node probing, exploit usage, or honeypot triggering.
    \item \textbf{Posterior Belief}:
    \begin{equation}\label{eq:bayesUpdate}
      \Pr(\theta \mid E_t) 
      \;=\; 
      \frac{\Pr(E_t \mid \theta)\,\Pr(\theta)}{\sum_{\theta' \in \Theta}\Pr(E_t \mid \theta')\,\Pr(\theta')},
    \end{equation}
    \item \textbf{Strategy Update}: Given updated beliefs, the defender recalculates optimal defense strategies for subsequent stages.
\end{enumerate}

This process refines the defender’s estimates about which attacker type is most likely, especially if different types use distinct routes or exploits. Observing that an attacker did not trigger a certain honeypot can shift probability mass away from types that would have been caught under those circumstances.

\subsection{Payoff Matrix}

In a single-round model, each cell of the payoff matrix yields a tuple \(\bigl(R_d,\; R_{a_1},\;\dots,\;R_{a_m}\bigr)\) when multiple attackers are involved. In multi-round scenarios, these expected utilities are recalculated whenever the defender updates \(\Pr(\theta \mid E)\).

For instance, if two attackers \(\bigl(a_{1}, a_{2}\bigr)\) exist and the defender has three pure strategies, the payoff matrix can be viewed as a \(3 \times 2 \times 2\) structure. One dimension corresponds to the defender’s choice \(\{\emptyset,\; \{v_1\},\; \{v_2\}\}\), while the other two dimensions reflect each attacker’s selected node \(\{v_1\}\) or \(\{v_2\}\). Each cell \(\bigl(d,\,(a_{1},a_{2})\bigr)\) stores a triple \(\bigl(R_d,\;R_{a_1},\;R_{a_2}\bigr)\). Table~\ref{tab:payoff-matrix} illustrates such a matrix with three defender actions and two actions for each attacker.

\begin{table}[h]
\centering
\caption{Example $3\times2\times2$ payoff matrix.}
\label{tab:payoff-matrix}
\renewcommand{\arraystretch}{0.92}
\setlength{\tabcolsep}{2pt}
\footnotesize
\resizebox{\linewidth}{!}{%
\begin{tabular}{c|cccc}
\toprule
& \multicolumn{4}{c}{\textbf{Attackers’ strategy}}\\
\cmidrule(lr){2-5}
& $(v_1,v_1)$ & $(v_1,v_2)$ & $(v_2,v_1)$ & $(v_2,v_2)$\\
\midrule
$d=\emptyset$   & $(R_d,R_{a1},R_{a2})$ & $(R_d,R_{a1},R_{a2})$
                & $(R_d,R_{a1},R_{a2})$ & $(R_d,R_{a1},R_{a2})$\\
$d=\{v_1\}$     & $(R_d,R_{a1},R_{a2})$ & $(R_d,R_{a1},R_{a2})$
                & $(R_d,R_{a1},R_{a2})$ & $(R_d,R_{a1},R_{a2})$\\
$d=\{v_2\}$     & $(R_d,R_{a1},R_{a2})$ & $(R_d,R_{a1},R_{a2})$
                & $(R_d,R_{a1},R_{a2})$ & $(R_d,R_{a1},R_{a2})$\\
\bottomrule
\end{tabular}}
\end{table}

\subsection{Expected Utility}

In a Bayesian setting, the attacker’s type \(\theta \in \Theta\) can affect which path \(s\) is optimal or how much damage the attacker might inflict. Similarly, the defender’s baseline reward might depend on \(\theta\) if different attacker types produce varying losses upon reaching certain nodes. Thus, the baseline reward functions are extended to
\[
R_d(x,\; s,\; \theta)
\quad\text{and}\quad
R_a(x,\; s,\; \theta),
\]
to capture attacker-type-specific costs, values, or behaviors. After observing events \(E\), the defender updates \(\Pr(\theta \mid E)\) according to \eqref{eq:bayesUpdate}.

To account for type uncertainty, the expected utilities become:
\[
U_{D}(x,\; s) 
\;=\; 
\sum_{\theta \in \Theta} \Pr\bigl(\theta \mid E\bigr)\; R_d\!\bigl(x,\;s,\;\theta\bigr),
\]
\\
\[
U_{a_i}(x,\; s) 
\;=\; 
\sum_{\theta \in \Theta} \Pr\bigl(\theta \mid E\bigr)\; R_a\!\bigl(x,\;s,\;\theta\bigr).
\]
Hence, \(R_d(x,s,\theta)\) and \(R_a(x,s,\theta)\) serve as baseline reward functions for a specific type \(\theta\), while \(U_{D}\) and \(U_{a_i}\) are the corresponding Bayesian-averaged rewards, each weighted by \(\Pr(\theta \mid E)\). This distinction allows the solver to pick an optimal defense strategy that anticipates how different attacker types might respond, and for each attacker to choose a path given its own type \(\theta\).

\subsection{Stackelberg Equilibrium for Multiple Followers}\label{subsec:DOBSS_multi}

To accommodate multiple followers simultaneously, we extend the original formulation to enforce a \emph{joint best-response} condition. Below, we provide a more detailed mathematical view of how this multi-follower extension can be encoded as a single MILP.
\begin{enumerate}
\item\textbf{Defender Mixed Strategy.}
Let \(\mathcal{D}\) be the defender’s set of pure strategies, where each \(d\in \mathcal{D}\) denotes, for instance, a particular subset of nodes at which honeypots are placed. We introduce variables
\[
  x_d \quad (\forall\,d\in \mathcal{D}),
\]
with the constraints
\[
  \sum_{d \in \mathcal{D}} x_d \;=\; 1,
  \quad
  x_d \;\ge\; 0,
\]
so that \(x_d\) is the probability of choosing defender pure strategy \(d\). This defines the defender’s \emph{mixed strategy} \(x\).

\item\textbf{Attacker Actions.}
Each attacker $A_i$ can choose from $\mathcal{A}_i$, where each $a_i \in \mathcal{A}_i$ denotes a feasible path with exploit combination. In a multi-follower scenario:

\[
  a_i \;\in\; \mathcal{A}_i,
  \quad
  \forall\,i=1,\dots,m.
\]

\item\textbf{ Reward Representation.}
We denote the defender’s reward when the defender plays \(d\) and attackers play \((a_1,\dots,a_m)\) by
\[
   R_{d}\bigl(d,\;a_1,\dots,a_m\bigr).
\]
Similarly, each attacker \(A_i\) has a reward
\[
   R_{a_i}\bigl(d,\;a_1,\dots,a_m\bigr).
\]
These rewards can be Bayesian-averaged if each attacker has uncertain types. For example, if \(\theta_i\) are possible types for attacker \(A_i\), then
\[
  R_{a_i}(d,a_1,\dots,a_m) \;=\;
  \sum_{\theta_i} \Pr(\theta_i)\;\widetilde{R}_{a_i}(d,a_1,\dots,a_m,\theta_i),
\]
and similarly for the defender’s reward.

\item\textbf{Joint Best Response.}
To capture each attacker’s best response—given that other attackers’ actions may also vary—we define indicator variables \(z_{(a_1,\dots,a_m)} \in \{0,1\}\). Specifically,
\[
   \sum_{(a_1,\dots,a_m)} z_{(a_1,\dots,a_m)} \;=\; 1,
\]
meaning exactly one joint action profile \((a_1,\dots,a_m)\) is “selected” by the attackers.
\end{enumerate}
Each attacker \(A_i\) must maximize its own expected utility \(\sum_d x_d\,R_{a_i}(d,a_1,\dots,a_m)\), subject to the actions \(\{a_j\}_{j \neq i}\) chosen by the other attackers. We can impose this via linear constraints of the form:

  For each attacker $A_i$, for each feasible profile $(a_1,\dots,a_m)$, \\
   \[
   \sum_{d \in \mathcal{D}} x_d \, R_{a_i}(d,\;a_1,\dots,a_m)
   \;+\;
   \beta \,\bigl(1 - z_{(a_1,\dots,a_m)}\bigr)\] \\ \vspace{-.7cm}
  \[ \;\;\ge\;\;\sum_{d \in \mathcal{D}} x_d \, R_{a_i}(d,\;a_1',\dots,a_m'),\]

for every alternative action \(a_i'\) attacker \(i\) might consider, when the other attackers’ actions are fixed at \((a_1,\dots,a_{i-1},a_{i+1},\dots,a_m)\). The constant \(\beta\) is a sufficiently large number.

In simpler words, whenever \(z_{(a_1,\dots,a_m)}=1\) indicates \((a_1,\dots,a_m)\) is chosen, attacker \(A_i\) must not have an alternative \(a_i'\) that yields a strictly higher reward. If \(z_{(a_1,\dots,a_m)}=0\), that profile is not chosen, and the constraint is automatically slack.

Hence the objective is
\[
\begin{aligned}
  &\max \; U_d,\\
& \text{subject to} \\
  & \;\;
  U_d \;\le\; \sum_{d\in \mathcal{D}} x_d \, R_{d}\bigl(d,\;a_1,\dots,a_m\bigr)
   \;+\; \beta\,\bigl(1 - z_{(a_1,\dots,a_m)}\bigr)
   \end{aligned}
\]
for every profile \((a_1,\dots,a_m)\), mirroring the single-follower DOBSS constraints but now extended to multiple simultaneous attackers.

By introducing the binary variables \(z_{(a_1,\dots,a_m)}\) for all attackers’ joint actions and coupling them through the best-response inequalities for each attacker, we obtain a single MILP that provides an optimal leader strategy \(x\). This solution anticipates all attackers’ actions under uncertainty, thereby extending the DOBSS framework to a multi-follower cybersecurity scenario.

\section{Implementation Methodology}\label{sec:Implementation}
In this section, we provide a detailed description of the implementation methodology used to construct and simulate our multi-attacker Bayesian Stackelberg security game. The structured approach outlined below serves as the basis for the concrete case study presented in Section~\ref{sec:case-study}. Algorithm~\ref{alg:MultiAttackerDOBSS} summarizes the entire implementation process.

\begin{algorithm}[H]
\caption{Multi-Attacker Bayesian Stackelberg}
\label{alg:MultiAttackerDOBSS}
\begin{algorithmic}[1]

\Procedure{INPUT}{$G,\mathcal{E},V,\kappa,\mathcal{T},C_h,\alpha,\Theta,P_0$}
  \State $V\leftarrow$ Compute node‐values$(G,\mathcal{T},\alpha)$
  \State $G_\theta\!\leftarrow$ Feasible Subgraphs$(G,\mathcal{E},\Theta)$
  \State $P\leftarrow P_0,\;U_d\leftarrow0,\;r\leftarrow1$
  \While{any attacker active}
    \State $E^{(r)}\leftarrow\mathrm{observe}()$
    \State $P\leftarrow\mathrm{BayesUpdate}(P,E^{(r)})$
    \State Build payoff $\{R_d,R_{a_i}\}$ for $d\in\mathcal{D},\,a_i\in\mathcal{A}_i(P)$
    \State Calculate Stackelberg equilibrium~$(x^*,\,\{a_i^*\})$
    \State {PlaceHoneypots}$(x^*)$
    \State $U_d = \mathbb{E}_{x^*}[R_d(x^*,a^*)]$
    \State Update Status()
    \State $r\leftarrow r+1$
  \EndWhile
  \State \Return $U_d$
\EndProcedure

\end{algorithmic}
\end{algorithm}

\subsection{Network Generation and Exploit Assignment}
We generate a directed graph representing the network, consisting of nodes (system components or resources) and directed edges (potential attacker traversal paths). Each edge is randomly assigned between one and four exploits, selected from a predefined set of ten distinct exploits. Every exploit carries a unique baseline cost , capturing the associated effort, resource consumption, or detection risk. By introducing variability in exploit assignments, attackers of differing skill levels encounter distinct path choices and traversal costs.

\subsection{Target Selection and Layered Structure}
Within the network, we strategically position multiple target nodes at varying depths, ensuring that attackers must navigate through several intermediate layers before reaching their objectives. This layered topology allows the defender multiple opportunities to intercept attackers or observe their behavior, enhancing the realism and effectiveness of the simulated scenarios.

\subsection{Attacker Skill and Path Choice}
Attackers are fully rational agents who prioritize minimizing the cumulative exploit cost of a path, then secondarily minimize the hop count, thereby arriving as early as possible. Three discrete skill categories (high, medium, low) define the exploit sets accessible to each attacker:
\begin{itemize}
  \item \textbf{High-skill:} Full exploit set, maximum path flexibility.
  \item \textbf{Medium-skill:} Moderate subset of exploits.
  \item \textbf{Low-skill:} Minimal subset, restricted to specific edges.
\end{itemize}
The feasible subgraph an attacker uses is determined by whether its skill set intersects the exploits on each edge.

\subsection{Bayesian Belief Updates}
Initially, we assume a uniform prior distribution over possible attacker types, capturing different skill and exploit profiles. As attackers move through the network, we observe which edges they traverse or whether they trigger honeypots. We update our posterior beliefs using Bayes’ rule:
\[
\Pr(\theta \,\mid\, E_t) 
\;=\; 
\frac{\Pr(E_t \,\mid\, \theta)\,\Pr(\theta)}{\sum_{\theta'} \Pr(E_t \,\mid\, \theta')\,\Pr(\theta')}.
\]
Any action inconsistent with a given type’s capabilities (e.g., using an exploit not in its set) reduces that type’s posterior probability, sharpening the defender’s understanding of the adversary’s skill profile.

\subsection{Stackelberg Equilibrium}
To determine defender and attacker strategies, we formulate a MILP following an extended approach:
\begin{enumerate}
  \item \textbf{Defender Strategies:} Enumerate all honeypot allocations over nodes, each allocation incurring a cost \(C_h\) per honeypot.
  \item \textbf{Attacker Actions:} Collect all feasible paths for each attacker, given exploit access.
  \item \textbf{Joint Best Response:} Introduce binary variables indicating which joint attacker profile is realized. Impose constraints so no attacker can gain higher utility by deviating from its chosen path.
  \item \textbf{Equilibrium Computation:} Solve the MILP to maximize the defender’s expected utility, yielding a mixed strategy for honeypot placement and best-response paths for each attacker.
\end{enumerate}

\subsection{Multi-Round Simulation}
This iterative process allows the defender’s honeypot deployments to adapt in response to newly inferred attacker capabilities, incrementally constraining attacker pathways.

\begin{figure}
    \centering
    \includegraphics[width= 0.9 \linewidth]{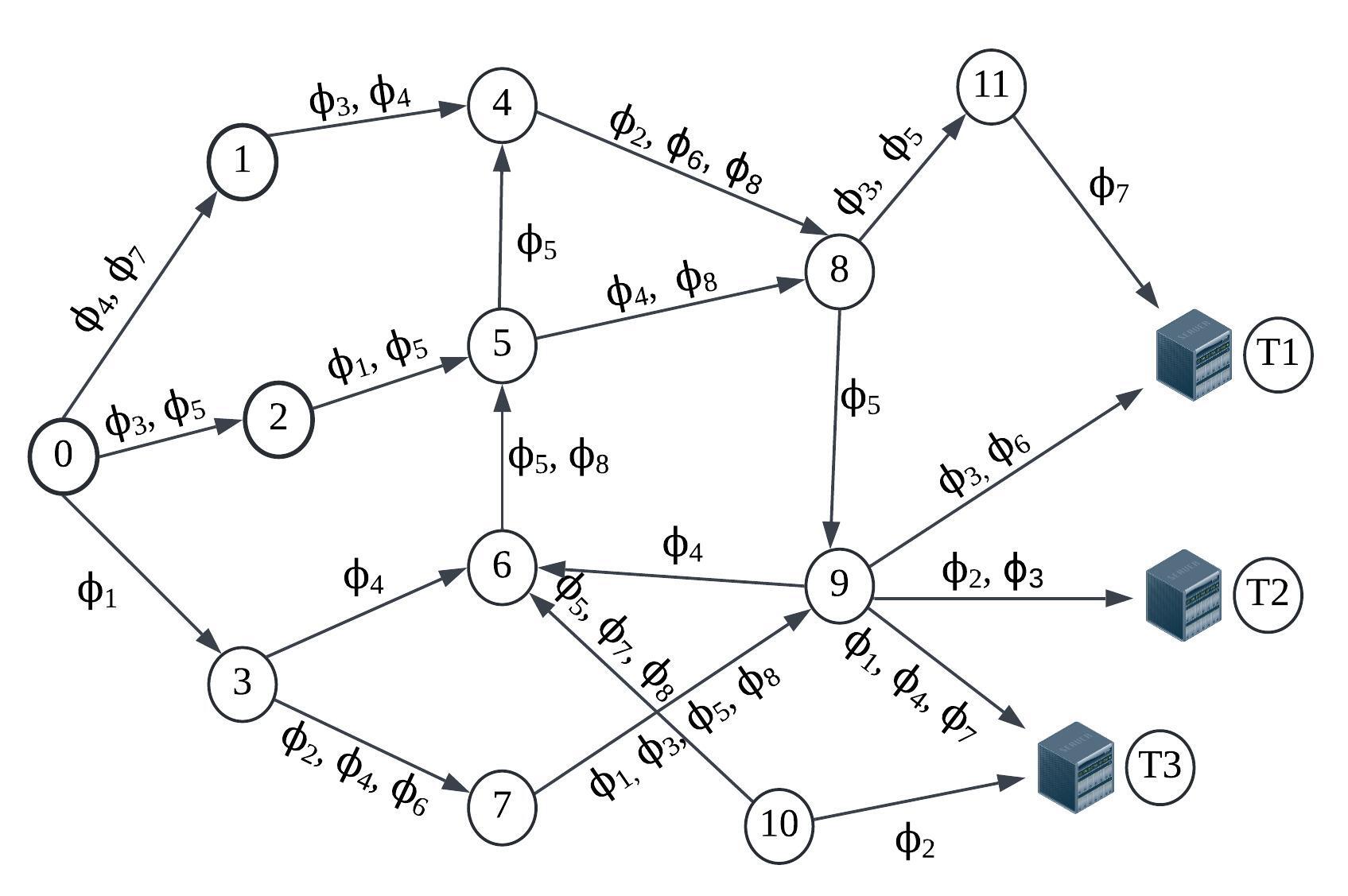} 
    \caption{Network Structure} 
    \label{fig:case_study_network} 
\end{figure}

\section{Case Study}
\label{sec:case-study}

This section illustrates our framework on a 15-node network comprising 21 directed edges, depicted in Figure~\ref{fig:case_study_network}. The discount factor is fixed at \(\alpha = 0.4\), and each honeypot deployment cost $C_{\mathrm{h}}= 3 $.
Each edge has one or more exploits that an attacker must possess to traverse it.  
We denote an exploit by $\phi_i$ and associate it with a cost $c_i$ that captures the time, effort, and detection risk incurred when that exploit is used.  
The eight exploits considered in this study are
\[
\begin{aligned}
(\phi_i, c_i) = \{\, &(\phi_1, 9.0),\; (\phi_2, 7.5),\;
(\phi_3, 2.5),\; (\phi_4, 4.0),\\
                  &(\phi_5, 3.0),\; (\phi_6, 1.5),\;
(\phi_7, 5.0),\; (\phi_8, 1.0)\,\}
\end{aligned}
\]

\begin{figure*}[!b]
  \centering
  \includegraphics[width=0.9\linewidth]{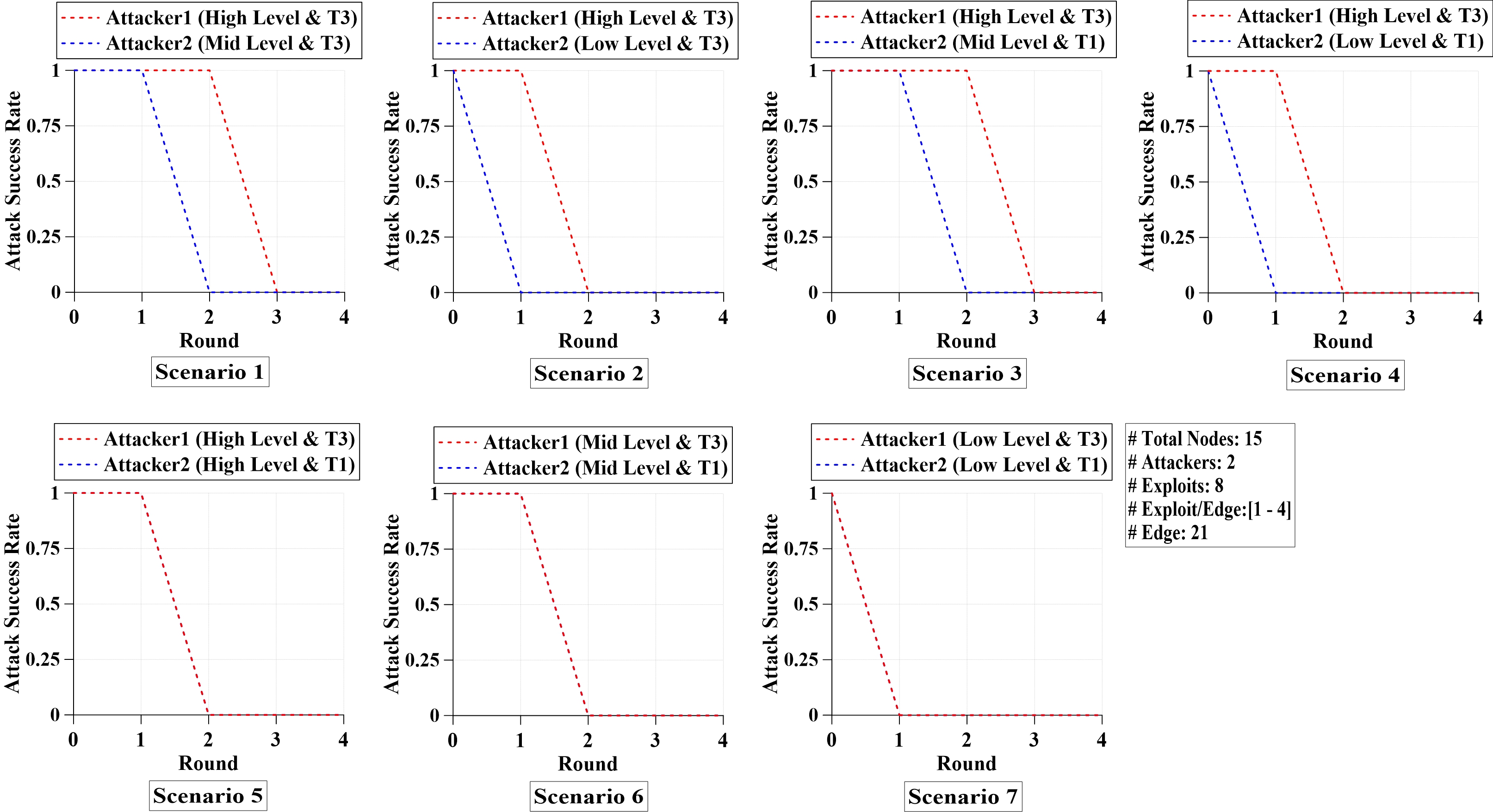}
  \caption{Attack Success Rate Across Rounds}
  \label{fig:scenarios}
\end{figure*}

Each attacker is characterized by the target node and by skill level that limits the exploits available to it.  
For instance, a Low-skill attacker may wield the set $\{\phi_3,\phi_5,\phi_6,\phi_8\}$, a Mid-skill attacker the set $\{\phi_3,\phi_4,\phi_5,\phi_6,\phi_7,\phi_8\}$, and a High-skill attacker the full set $\{\phi_1,\dots,\phi_8\}$.  
An edge that requires exploits $\{\phi_4,\phi_7\}$ can be traversed only by attackers holding at least one of those exploits.  
Consequently, each attacker first constructs a feasible subgraph that contains only the edges it can penetrate and then selects a path from its start node to a target node that minimizes the cumulative exploit cost:
\[
\textstyle \sum_{e\in P} c(\phi_e),
\]
where $P$ is the chosen path and $\phi_e$ is the cheapest exploit that unlocks edge~$e$ for that attacker.

In Figure~\ref{fig:case_study_network}, node~$0$ is the common entry point for all attackers, whereas nodes $T_1$, $T_2$, and $T_3$ are targets with values of $15$, $10$, and $20$, respectively.  
After identifying its feasible subgraph, an attacker computes the minimum-cost route from node~$0$ to one of the targets.  
Meanwhile, the defender observes the attackers’ moves, calculates a Stackelberg Equilibrium to determine optimal honeypot placements, and adaptively deploys or removes honeypots to maximize expected utility.

Figure~\ref{fig:scenarios} presents the attack success rates over five discrete rounds (0-4) for seven scenarios, each combining two attackers of varying skill levels (High, Mid, or  Low) and aiming at either the same or different targets ($T_3$ and/or $T_1$). The vertical axis shows the attackers’ success rate at penetrating the network in each round. We observe the following:

\textbf{Scenario 1: (High vs Mid), same target $T_3$.}
The High-skill attacker succeeds in rounds 1 and 2, but fails from round 3 onward.
The Mid-skill attacker succeeds in round 1, yet is stopped starting in round 2.
Thus honeypots block the Mid attacker one round earlier, while the High attacker’s broader exploit set keeps a viable path open slightly longer.

\textbf{Scenario 2: (High vs Low), same target $T_3$.}
The High attacker penetrates in round 1, then fails at round 2.
The Low attacker is blocked from round 1.
Limited exploits make the Low attacker easy to neutralize, whereas the High attacker forces an extra round of defensive adjustment.

\textbf{Scenario 3: (High vs Mid), different targets $T_3$ and $T_1$.}
The High attacker succeeds through round 2 before being stopped at round 3.
The Mid attacker succeeds in round 1 and fails from round 2.
Separate targets initially give both attackers room to maneuver, yet the High attacker’s additional exploits extend its run by one round.

\textbf{Scenario 4: (High vs Low), different targets $T_3$ and $T_1$.}
The High attacker succeeds in round 1, then is intercepted at round 2.
The Low attacker fails from round 1.
Even with targets split, the Low attacker’s limited exploit set leads to an early shutdown; the High attacker forces the defender to invest in more honeypots before failing.

\textbf{Scenario 5: (High vs High), different targets $T_3$ and $T_1$.}
Both High attackers reach their goals in round 1, but neither succeeds in round 2 or beyond.
Full exploit sets allow swift initial penetration, yet the defender adapts quickly enough to prevent repeated breaches.

\textbf{Scenario 6: (Mid vs Mid), different targets $T_3$ and $T_1$.}
Both Mid attackers fail from round 2 onward.
Once honeypots are aligned with their paths, their intermediate skill no longer suffices to evade detection.

\textbf{Scenario 7: (Low vs Low), different targets $T_3$ and $T_1$.}
After the first move, the defender’s revised honeypot layout fully neutralizes their limited exploit options.

The results exhibit a clear pattern across all scenarios. Attackers who succeed initially lose that edge after one or two rounds. High-skill attackers typically survive longer than Mid-skill ones due to having more options, but they too are neutralized once the defender adapts. Low-skill attackers rarely maintain success in round~1, largely because they have fewer exploits to circumvent newly placed honeypots. Overall, the data show that, although attackers exploit the network’s initial vulnerabilities, the defender’s dynamic honeypot strategy rapidly reduces attack success rates to zero in subsequent rounds.


\section{Experimental Results and Analysis}
\label{sec:Experimental_Evaluation}

This section discusses the environment of our experiments and the performance of our proposed framework in a two-attacker scenario under varying conditions. 
\subsection{Experimental Setup}
In this experiment, we generate a directed network of \(n = 100\) nodes, ensuring at least \(3n\) edges to maintain a rich variety of paths. We place three distinct targets within the network, each lying at a BFS depth of at least six from the main entry node. These targets are assigned relatively high values (e.g., 10.0, 15.0, and 20.0) to reflect the damage inflicted upon the defender if compromised. Requiring nodes at BFS depth six ensures that attackers must move through multiple intermediate layers, giving the defender opportunities to intercept or observe their actions. Throughout all experiments, we fix the basic discount factor at \(\alpha = 0.4\) and set the honeypot deployment cost to \(C_h = 3\). We assume that all the attackers have high-skill. In particular, we examine how the basic discount factor $\alpha$ influences defender utility, compare our method to simpler baseline strategies, and then analyze the computational scalability of our approach as the network grows larger. All tests seek to demonstrate that our method consistently identifies beneficial honeypot allocations while remaining tractable for increasing problem sizes.

\subsection{Sensitivity Analysis}
\label{subsec:sensitivity}

\subsubsection{Sensitivity w.r.t. Discount Factor ($\alpha$)}

\begin{figure}
    \centering
    \includegraphics[width= 0.85 \linewidth]{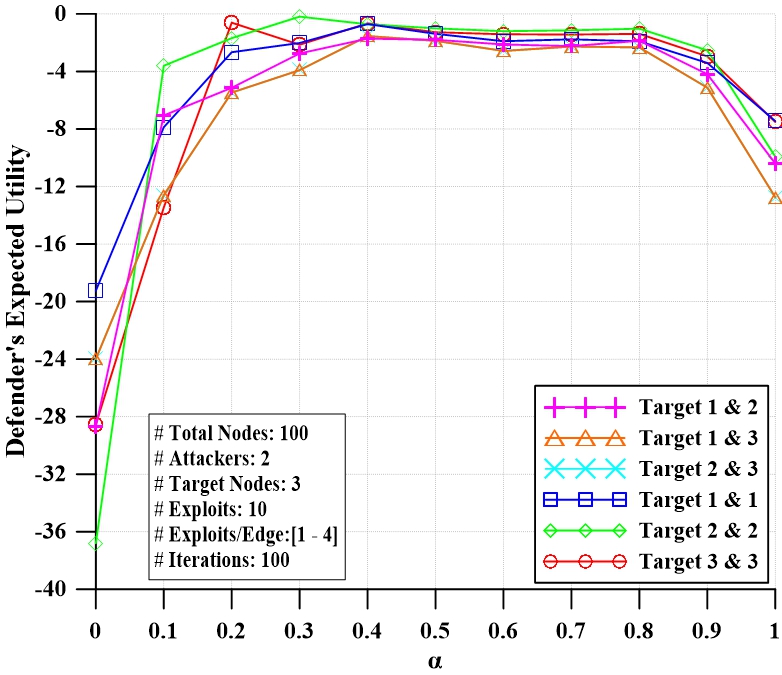}
    \caption{Impact of the discount factor}
    \label{fig:alpha_impact}
\end{figure}

The discount factor $\alpha$ appears in the node-value computation (Section~\ref{sec:Implementation}) and determines how quickly non-target nodes lose their worth as the distance to a target grows. As shown in Figure~\ref{fig:alpha_impact}, each curve corresponds to a distinct pair of target nodes pursued by the two attackers. We observe that when $\alpha$ is small, in the range $[0,0.3]$, the values of non-target nodes become so insignificant that the optimization procedure deems it preferable not to place any honeypots.
As $\alpha$ increases to the intermediate band around $[0.4,0.8]$, the discounted node values strike a balance that motivates the  defender to protect mid-depth nodes without incurring excessive costs, and the resulting utility stays relatively constant. Finally, as $\alpha$ exceeds 0.8, nearly every node in the network becomes as valuable as, or even more valuable than, the designated targets. In such a high-valuation regime, a single path that bypasses a honeypot can seriously erode the defender’s payoff, pushing the utility downward again as $\alpha$ approaches  1.
\begin{figure}
    \centering
    \includegraphics[width= 0.85 \linewidth]{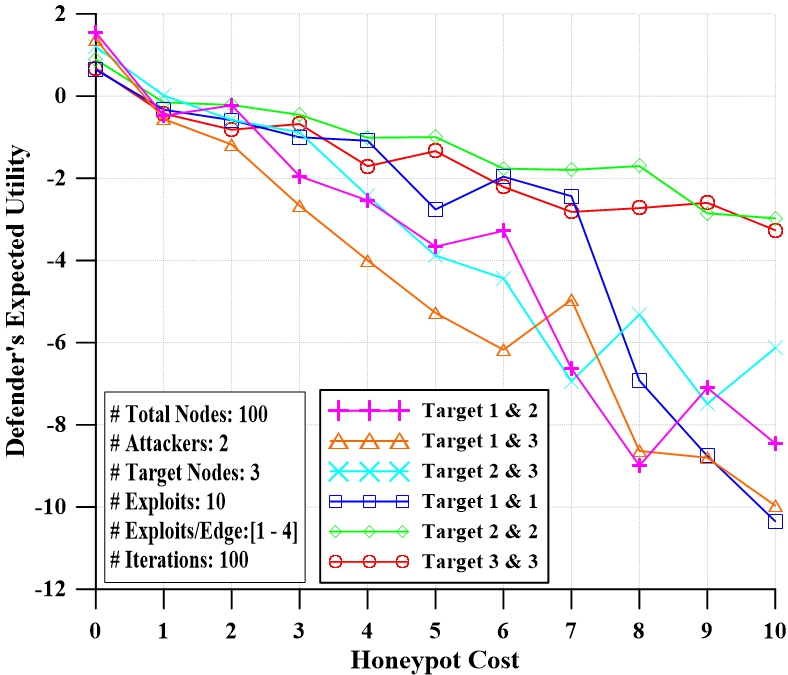}
    \caption{Impact of the honeypot cost}
    \label{fig:honeypot_cost_sensitivity}
\end{figure}

\subsubsection{Sensitivity w.r.t. Honeypot Cost ($C_{\mathrm{h}}$)}

Figure~\ref{fig:honeypot_cost_sensitivity} shows how the defender’s expected utility declines as honeypot cost rises from 0 to 10, considering six target-node pairings. When costs remain below 3, different attacker strategies yield similar utility levels. After cost 3, strategies involving distinct targets begin to diverge more sharply from same-target strategies. This gap continues to widen until around cost 7, at which point even the same-target approach converges toward the utility of different-target scenarios. These results indicate that moderate honeypot costs underscore the advantages or disadvantages of overlapping attacker objectives, while very high costs tend to compress the difference in defender utilities. Overall, increasing cost forces the defender to be more selective with honeypot placement, emphasizing the importance of a cost-aware allocation strategy that balances coverage with potential trade-offs in utility.
\begin{figure*}[!b]
    \centering
    \includegraphics[width= 0.8 \linewidth]{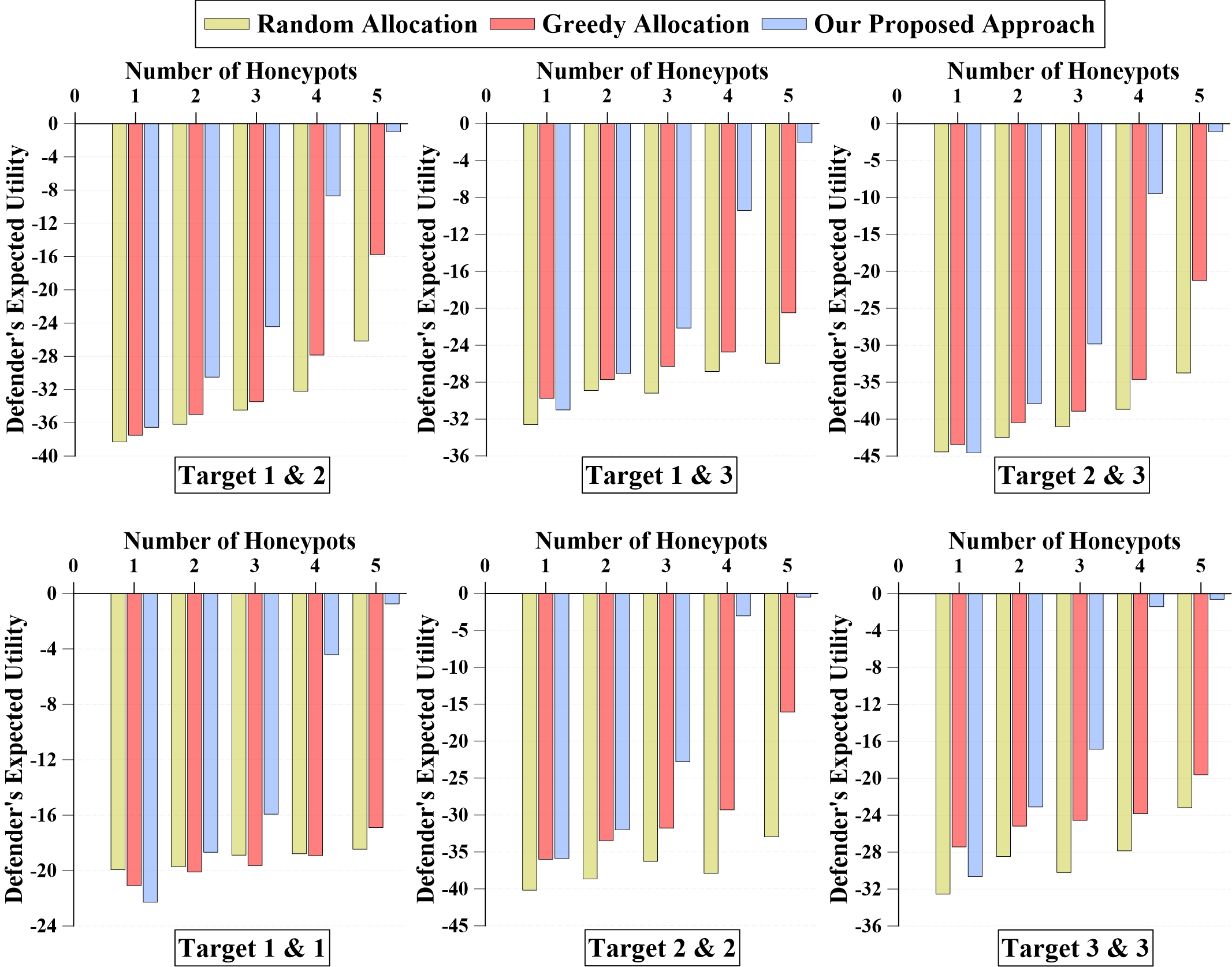} 
    \caption{Defender's Expected Utility versus the number of honeypots} 
    \label{fig:defender_utility_comparison} 
\end{figure*}
\subsection{Utility Analysis}
\label{subsec:baselines}

To evaluate how effectively our approach manages the defense of critical nodes, we compare it against two simpler methods, which are the “greedy” allocation of honeypots on the highest-valued nodes and a purely ``random” allocation. The defender can allocate between 0 and 5 honeypots among the nodes in each round. At the outset, our approach assumes a uniform prior probability over all possible attacker types and solves for a Stackelberg equilibrium strategy using this initial belief. After each round, the experiment updates the defender’s belief distribution based on the observed exploits, refines the Stackelberg solution, and then repeats until the attackers are caught or reach their respective targets. Figure~\ref{fig:defender_utility_comparison} indicates that when only one honeypot is deployed, our proposed method can occasionally produce a lower defender utility than the other two strategies. This outcome arises because the Bayesian probability updates in the \emph{first} round are initialized uniformly; under certain attacker configurations, placing a single honeypot at an equilibrium location may not protect critical paths as effectively from the outset. However, as the game progresses through the second and third rounds, updated exploit information and repeated Stackelberg planning enable the method to intercept the most threatening routes more reliably. Consequently, the defender’s utility tends to increase significantly in later rounds, outstripping both the greedy strategy and the random strategy.

\subsection{Scalability Analysis}

\label{subsec:scalability}

\begin{figure}[t]
  \centering
  \includegraphics[width=0.9\linewidth]{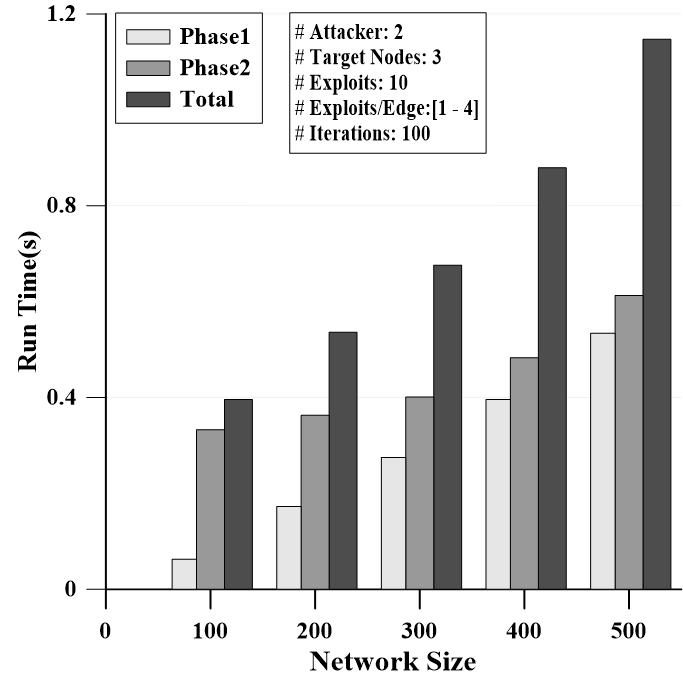}
  \caption{Runtime Across Network Sizes}
  \label{fig:scalability_chart}
\end{figure}

To validate scalability, we conduct experiments on networks of increasing size, from $100$ to $500$ nodes. Each network is constructed with at least $3n$ directed edges. Concurrently, we increase the BFS distance for target nodes beyond 6, so that the attackers have to traverse progressively longer paths. We divide the runtime into two phases: Phase 1 payoff-matrix construction and Phase 2 Stackelberg-equilibrium computation.
In this step, we enumerate possible defender allocations and attacker path choices to build the payoff matrix. This process includes identifying all feasible exploits per edge, computing node valuations, and determining potential honeypot positions. The chart’s light-gray bars in Figure~\ref{fig:scalability_chart} show how Phase1 time grows moderately with network size.
Once the payoff matrix is set, we solve the MILP to find equilibrium strategies for the defender and attackers. The dark-gray bars in Figure~\ref{fig:scalability_chart} indicate that Phase2 grows somewhat faster than Phase1 due to the complexity of joint best-response constraints in multi-attacker scenarios. 
The combined run time (the black bars) demonstrates that our approach maintains a practical solution scale by separating payoff construction from equilibrium solving. These findings illustrate that layering the network, focusing on feasible exploit sets, and maintaining a compact payoff representation keep the computational overhead manageable for sizable problems.

Overall, the experiments confirm that our proposed approach can handle networks with hundreds of nodes, multiple target nodes, and extensive attack surfaces. By rapidly identifying honeypot placements that intercept adversaries in critical areas, the method effectively scales to larger systems.

\subsection{Discussion and Limitations}
\label{subsec:limitations}
The presented Bayesian Stackelberg framework integrates multi-attacker dynamics, belief updating, and a game-theoretic resource-allocation strategy for honeypot placement. Its iterative BFS-layered approach enables the defender to monitor adversaries’ step-by-step progression, updating strategy accordingly. In each round, the MILP formulates ``joint best-response'' constraints for multiple attackers, offering an equilibrium-based solution.

Certain limitations follow from the assumptions and modeling abstractions. First, the approach assumes that all the attackers maintain rational objectives, high skill level, and stable motivations, which may not be valid if attackers rapidly adapt or if their goals evolve in response to partial information. Second, even though our scalability results suggest that networks of a few hundred nodes can be tackled within reasonable time, significantly, larger or more complex real-world infrastructures with a richer variety of attacker types may require further approximation or specialized heuristics to manage the combinatorial explosion. Third, real-world cyberdefense often involves factors like zero-day exploits, partial observability, or heterogeneous security teams, none of which are explicitly modeled here. Introducing these complexities could further challenge the belief-updating process and the defender’s capacity for swift honeypot redeployment. Finally, while the discount factor $\alpha$ meaningfully captures diminishing node value with distance, some networks may involve time-dependent or priority-based valuations, necessitating more dynamic or contextual metrics in the optimization framework.

Nevertheless, the approach demonstrates how an equilibrium-based defense strategy can effectively guide limited honeypot placements and belief revisions about attackers’ objectives. By leveraging a Bayesian Stackelberg formulation, the defender can anticipate key attacker behaviors and optimize defensive measures accordingly.

\section{Conclusion and Future Work}
\label{sec:conclusion}

This paper introduced a multi-attacker Bayesian Stackelberg game model that strategically integrates honeypot deployment, exploit-aware path analysis, and iterative belief updates. Experimental evaluations indicate that this approach not only surpasses traditional baselines such as greedy or random honeypot placement but also scales effectively to networks of up to 500 nodes with more than 1,500 edges within near-second total run times. By dynamically reallocating honeypots and refining attacker-type beliefs after each observation, the defender anticipates adversaries’ best responses and substantially reduces their success rates over successive rounds. These findings underscore the value of modeling cyber defense as a Bayesian Stackelberg game, allowing systematic adaptation to varying threat environments and efficient protection of critical assets.

Future research directions include investigating coordinated attackers who share intelligence, further enhancing the realism of adversarial strategies. Incorporating time-sensitive or event-driven reconfiguration of honeypot deployments can also increase the defender’s responsiveness to emerging threats. Broadening the model to account for zero-day exploits, partial observability, and heterogeneous security measures would offer further insight into defending large-scale, dynamic networks.

\bibliographystyle{IEEEtran}
\bibliography{References}
\end{document}